\documentclass{PoS}
\usepackage{amsmath,amssymb}

\newcommand{\be}{\begin{equation}}
\newcommand{\ee}{\end{equation}}
\newcommand{\bea}{\begin{eqnarray}}
\newcommand{\eea}{\end{eqnarray}}
\newcommand{\beq}{\begin{equation}}
\newcommand{\eeq}{\end{equation}}
\newcommand{\beqa}{\begin{eqnarray}}
\newcommand{\eeqa}{\end{eqnarray}}
\newcommand{\ba}{\begin{array}}
\newcommand{\ea}{\end{array}}

\title{{\em U}(2) and MFV in Supersymmetry}
\ShortTitle{$U(2)$ and MFV in Supersymmetry}

\author{\speaker{David M. Straub}
\\
        Scuola Normale Superiore and INFN, Piazza dei Cavalieri 7, 56126 Pisa, Italy\\
        E-mail: \email{david.straub@sns.it}}

\abstract{%
A suitably broken $U(2)^3$ symmetry acting on the first two generations of quarks and squarks
is studied as a natural generalization of the Minimal Flavour Violation principle to the case of a hierarchical squark spectrum.
A definite correlation emerges between the 
meson mixing amplitudes in the $K$, $B_d$ and $B_s$ systems which can resolve the current tension 
in CKM fits,
while predicting a moderate enhancement of the $B_s$ mixing phase, in agreement with recent data from the LHCb experiment, as well as sbottom and gluino masses in the TeV region.
}

\FullConference{The 2011 Europhysics Conference on High Energy Physics, EPS-HEP 2011,\\
		July 21-27, 2011\\
		Grenoble, Rh\^one-Alpes, France}

\begin{document}

\section{Setup}

\noindent
Supersymmetry with a hierarchical sfermion spectrum, {\em i.e.} with light third generation sfermions and multi-TeV masses for the first two generation ones,
allows to solve the SUSY CP problem without sacrificing the SUSY solution to the gauge hierarchy problem \cite{Cohen:1996vb,Barbieri:2010pd,Barbieri:2010ar,Barbieri:2011vn} because the experimentally accessible electric dipole moments (EDMs) are related to first generation fermions and are strongly suppressed if the corresponding superpartners are heavy.
Moreover, such spectrum is favoured by LHC sparticle searches: while the bounds on first generation squark masses are approaching a TeV, the third generation ones can still be significantly lighter \cite{ATLAS-CONF-2011-098}.
On the other hand, a hierarchical spectrum is not enough to suppress flavour violation for a generic soft SUSY breaking sector \cite{Giudice:2008uk}. Imposing Minimal Flavour Violation (MFV) is not an option either, since the $U(3)^3$ symmetry assumed to be only weakly broken by the Yukawa couplings is explicitly broken to $U(2)^3$ already by the squark hierarchy.
A $U(2)$ symmetry in turn has been used to explain, at least in part, the hierarchies manifest in the Yukawa couplings \cite{Pomarol:1995xc,Barbieri:1995uv}. However, a single $U(2)$ acting on left- and right-handed fields turns out not to provide enough protection from flavour violation.

Motivated by these observations, a $U(2)^3$ symmetry has been considered in \cite{Barbieri:2011ci} together with an appropriate breaking pattern as an alternative to MFV. The starting point is the global flavour symmetry $G_f=U(2)_{Q_L}\times U(2)_{U_R}\times U(2)_{D_R}$. Analogously to the MFV case, a pair of bi-doublet spurions is introduced, transforming as $\Delta Y_u= (2, \bar{2}, 1)$ and $\Delta Y_d= (2, 1, \bar{2})$.
To allow for communication between the third generation and the first two, at least one additional spurion is needed. The minimal choice compatible with the observed quark masses and mixings is a doublet transforming as $V = (2,1,1)$.
This setup has interesting consequences for CP violation in meson mixing.

\section{Consequences for meson mixing}

\noindent
A model-independent prediction of the $U(2)^3$ symmetry and its minimal breaking pattern is a {\em universal} modification of the $B_d$ and $B_s$ mixing amplitudes with a possible new phase as well as a contribution to the neutral kaon mixing amplitude which is aligned in phase with the SM.
In the particular case of SUSY with hierarchical sfermions, these contributions arise dominantly from gluino-sbottom box diagrams.  They modify
the CP violating parameter in the $K$ system $\epsilon_K$, the mass differences in 
the $B_{d,s}$ systems $\Delta M_{d,s}$ and the mixing induced CP 
asymmetries in $B_d\to\psi K_S$
and $B_s\to\psi\phi$
in the following way,
\begin{align}
\epsilon_K&=\epsilon_K^\text{SM(tt)}\times\left(1+x^2F_0\right) +\epsilon_K^\text{SM(tc+cc)},
\label{eq:epsKxF}\\
S_{\psi K_S} &=\sin\left(2\beta + \text{arg}\left(1+xF_0 e^{2i\gamma_L}\right)\right) , &
\Delta M_d &=\Delta M_d^\text{SM}\times\left|1+xF_0 e^{2i\gamma_L}\right| ,
\label{eq:DMdxF}\\
S_{\psi\phi} &=\sin\left(2|\beta_s| - \text{arg}\left(1+xF_0 e^{2i\gamma_L}\right)\right) ,
&
{\Delta M_d}/{\Delta M_s} &= {\Delta M_d^\text{SM}}/{\Delta M_s^\text{SM}} ,
\label{eq:MdMs}
\end{align}
where $x$ is a free $O(1)$ ratio of mixing angles (chosen to be positive without loss of generality) and $\gamma_L$ a free phase. $F_0$ is a positive loop function depending on the masses of the dominantly left-handed sbottom squark and the gluino (see \cite{Barbieri:2011ci}).

This pattern of effects in $\Delta F=2$ observables is
particularly interesting in view of
recently mounting tensions 
in the CKM description of CP violation, mostly among
$|\epsilon_K|$, $\Delta M_d/\Delta M_s$ and $S_{\psi K_S}$, which in the SM measures $\sin(2\beta)$ 
\cite{Lunghi:2008aa,Buras:2008nn,Altmannshofer:2009ne,Lunghi:2010gv,Bevan:2010gi}.
This tension, if explained by new physics, could be due to non-standard contributions in neutral kaon mixing, non-standard CP violation in $B_d$ mixing or a non-universal modification of the mass differences in the $B_d$ and $B_s$ systems. In $U(2)^3$, a combination of the first two possibilities is realized. The solution to the tensions in turn entails a prediction for CP violation in $B_s$ mixing.
To quantify these statements, we performed a global fit of the four CKM parameters as well as the model parameters $x$, $\gamma_L$ and $F_0$ to all the relevant $\Delta F=2$ constraints, taking into account the new physics contributions in (\ref{eq:epsKxF}--\ref{eq:MdMs}).
In the following, the results are presented with and without the inclusion in the fit of the recent LHCb result on the $B_s$ mixing phase \cite{LHCb-CONF-2011-056}.\footnote{At the time of presenting this talk, a large $B_s$ mixing phase was indicated by the data \cite{Lenz:2010gu}.
As the following plots show, the range predicted for $S_{\psi\phi}$ in the fit without the $B_s$ data
agrees well with the recent results by LHCb (using the average of $B_s\to\psi\phi$ and $B_s\to\psi f_0$ data).
}

\begin{figure}[tbp]
\begin{center}
\includegraphics[width=0.32\textwidth]{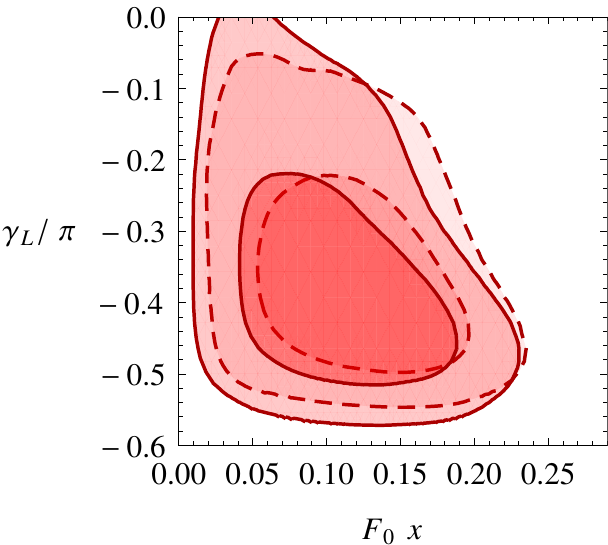}
\includegraphics[width=0.32\textwidth]{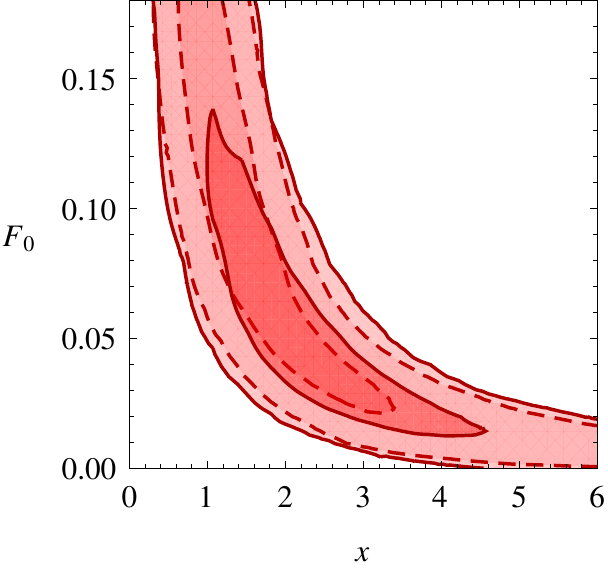}
\includegraphics[width=0.32\textwidth]{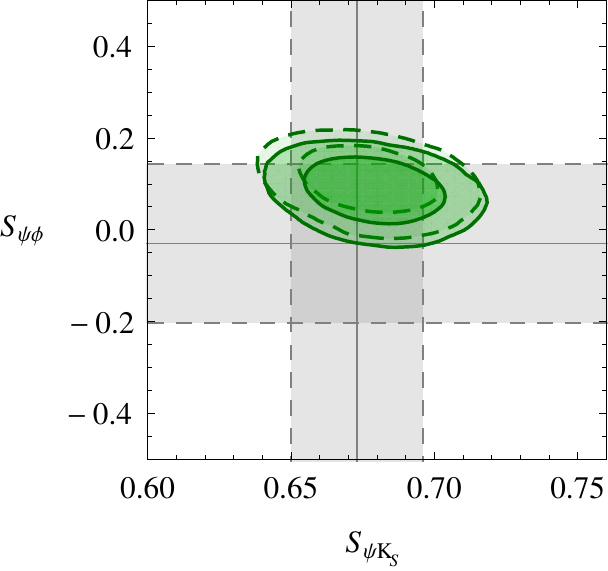}
\end{center}
\caption{%
Left and center: Preferred values of the parameters defined in the text, as determined from the fit. The dashed contours correspond to the 68\% and 90\% C.L. regions in the fit without $S_{\psi\phi}$, the solid contours to the fit including the new LHCb data.
Right: Fit results for $S_{\psi\phi}$ and $S_{\psi K_S}$ as determined from the fit. The gray bands are the experimental $1\sigma$ ranges.
In all plots, the dashed contours correspond to the 68\% and 90\% C.L. regions in the fit without $S_{\psi\phi}$, the solid contours to the fit including the new LHCb data.}
\label{fig:u2fit}
\end{figure}

The preferred ranges for the $U(2)^3$ parameters are shown in figure~\ref{fig:u2fit}.
The left panel shows the preferred values for the combination $F_0 x$ and the phase $\gamma_L$ entering $B_{d,s}$ mixing.
As shown in the center panel, $F_0$ and $x$ are not very well constrained separately, but $F_0\gtrsim0.05$ is preferred by the fit, implying sub-TeV gluino and squark masses.
The non-zero value of $\gamma_L$ required by the fit to solve the CKM tensions implies non-standard CP-violation in the $B_s$ system by means of equation~(\ref{eq:MdMs}).
In the right panel of figure~\ref{fig:u2fit}, we show the fit prediction for $S_{\psi\phi}$  in the $S_{\psi K_S}$ vs.~$S_{\psi\phi}$ plane in the fit without $S_{\psi\phi}$ as well as the result of the fit including both observables in the fit, using the new LHCb data. The results show that the CKM fit prefers a moderate positive value of $S_{\psi\phi}$, which is compatible with the LHCb results.

\section{Conclusion and outlook}

\noindent
As an alternative to MFV, a $U(2)^3$ symmetry makes model-independent predictions for CP violation in meson mixing. In particular, assuming the current tensions in the CKM fit to be resolved by new physics, it predicts a moderate enhancement of the $B_s$ mixing phase compatible with recent LHCb data. In the context of SUSY with hierarchical squark masses, this points to sbottom and gluino masses below about 1~TeV.
Imminent improvements of LHC bounds on 3rd generation squark masses as well as the $B_s$ mixing phase will allow a test of these predictions.
If realized, visible effects are also possible in CP asymmetries in $\Delta B=1$ decays, as is discussed in \cite{Barbieri:2011fc}.

\paragraph{Acknowledgments}
I thank R. Barbieri, G. Isidori, J. Jones-P\'erez and P. Lodone for the pleasant collaboration on the topic presented in this talk, P. Campli and F.~Sala for useful discussions and F.~Sala for comments on the manuscript.
This work was supported by the EU ITN ``Unification in the LHC Era'', 
contract PITN-GA-2009-237920 (UNILHC).

\bibliographystyle{JHEP-2}
\bibliography{straub}

\end{document}